\documentclass[fleqn,usenatbib]{mnras}
\usepackage{arydshln}
\usepackage{amssymb}
\usepackage{amsmath}% Extra maths symbols
\usepackage{graphicx}	% Including figure files

\title{Detecting the neutral IGM in filaments with the SKA}
\author[Kooistra et al.]{Robin Kooistra$^{1,2}$,
Marta B. Silva$^{2,3}$,
Saleem Zaroubi$^{2,4,5}$,
Marc A. W. Verheijen$^{2,6}$\newauthor
Elmo Tempel$^{7,8}$
and Kelley M. Hess$^{2,9}$
\\
$^1$Kavli IPMU (WPI), UTIAS, The University of Tokyo, Kashiwa, Chiba 277-8583, Japan\\
$^2$Kapteyn Astronomical Institute, University of Groningen, Landleven 12, 9747 AD Groningen, the Netherlands\\
$^3$Institute of Theoretical Astrophysics, University of Oslo,P.O. Box 1029 Blindern, N-0315 Oslo, Norway\\
$^4$Department of Natural Sciences, Open University of Israel, 1 University Road, PO Box 808, Ra'anana 4353701, Israel\\
$^5$Department of Physics, The Technion, Haifa 32000, Israel\\
$^6$Adjunct Faculty, National Centre for Radio Astrophysics, TIFR, Ganeshkhind, Pune 411007, India\\
$^7$Leibniz-Institut f\"{u}r Astrophysik Potsdam (AIP), An der Sternwarte 16, 14482 Potsdam, Germany\\
$^8$Tartu Observatory, University of Tartu, Observatooriumi 1, 61602 T\~{o}ravere, Estonia\\
$^9$ASTRON, Netherlands Institute for Radio Astronomy, PO 2, 7990 AA, Dwingeloo, The Netherlands}

\begin{document}
\maketitle

\begin{abstract}
The intergalactic medium (IGM) plays an important role in the formation and evolution of galaxies. Recent developments in upcoming radio telescopes are starting to open up the possibility of making a first direct detection of the 21 cm signal of neutral hydrogen (HI) from the warm gas of the IGM in large-scale filaments. The cosmological hydrodynamical EAGLE simulation is used to estimate the typical IGM filament signal. Assuming the same average signal for all filaments, a prediction  is made for the detectability of such a signal with the upcoming mid-frequency array of the Square Kilometer Array (SKA1-mid) or the future upgrade to SKA2. The signal-to-noise (S/N) then only depends on the size and orientation of each filament. With filament spines inferred from existing galaxy surveys as a proxy for typical real filaments, we find hundreds of filaments in the region of the sky accessible to the SKA that can be detected. Once the various phases of the SKA telescope become operational, their own surveys will be able to find the galaxies required to infer the position of even more filaments within the survey area. We find that in 120 h, SKA1-mid/SKA2 will detect HI emission from the strongest filaments in the field with a S/N of the order of 10 to $\sim150$ for the most pessimistic model considered here. Some of the brighter filaments can be detected with an integration time of a few minutes with SKA1-mid and a few seconds with SKA2. Therefore, SKA2 will be capable of not only detecting but also mapping a large part of the IGM in these filaments.
\end{abstract}


\defcitealias{art:kooistra17}{K17}
\defcitealias{art:hm18}{P18}
\defcitealias{art:hm12}{HM12}
\defcitealias{art:hm01}{HM01}

\section{Introduction}\label{sec:intro}
In the standard picture, the structure in the Universe forms through non-linear gravitational collapse. This creates an intricate pattern of galaxies, filaments and voids, collectively known as the cosmic web. Such structures are clearly seen in simulations based on the $\Lambda$-Cold Dark Matter ($\Lambda$CDM) cosmological model.  Simulations can provide information on both the dark matter and the baryonic gas particles, whereas, on the observational side, the main probe of the large-scale structure has been through the three-dimensional distribution of the observed galaxies, such as in the Sloan Digital Sky Survey \citep[SDSS, ][]{art:sdss1,art:sdss8}, the 6-degree Field (6dF) Galaxy Redshift Survey \citep{art:6df1,art:6df3} and the Two Micron All-Sky Redshift Survey \citep[2MRS, ][]{art:2mrs}. Specifically, a significant effort has been made in inferring and characterizing large-scale filaments from galaxy surveys and they seem to match the predictions made by the $\Lambda$CDM model \citep[e.g.,][]{art:sousbie08,art:jasche10,art:smith12,art:tempel14}. Unfortunately however, the positions of galaxies only provide a biased tracer of the underlying density field and give little information on the gas content in the filaments themselves.\\

\noindent Recent observations have begun to probe the hydrogen gas from the intergalactic medium (IGM). In particular, cross-correlation of neutral hydrogen (HI) 21 cm intensity maps, together with galaxy surveys at $z\sim0.8$ have given strong evidence for the existence of HI gas in galaxies below the detection limit in the cosmic web \citep{art:chang10,art:masui13}. Furthermore, at higher redshift, signs of filamentary structure in the gas have been detected in absorption in the spectra of background sources \citep[e.g.,][]{art:abs1,art:abs2,art:abs3,art:finley14}. Nonetheless, the number of direct detections of the IGM in filaments is still limited to a few in the vicinity of galaxies in the circumgalactic medium and the size of the detected filaments is relatively small at a few $\sim100$ kpc \citep[e.g.,][]{art:film31m33,art:fildet}.\\

\noindent The IGM can be divided into two main components: the Warm-Hot Intergalactic Medium (WHIM) with temperatures of $\sim10^5-10^7$ K and a cooler component with temperatures up to $\sim10^5$ K ~\citep{art:cen93,art:evrard94}. The WHIM consists of highly ionized gas that is heated by local galaxies, but predominantly shock-heated during structure formation, and is expected to contain a large fraction of the baryonic matter in the IGM ~\citep[e.g.,][]{art:yoshida05}. The existence of significant amounts of hot gas in large-scale filaments has been confirmed through X-ray observations ~\citep[e.g.,][]{art:eckert15}, by tracing the galaxy luminosity density ~\citep{art:nevalainen15} and through the thermal Sunyaev-Zel'dovich effect ~\citep{art:szeffect}. The cooler component of the IGM, on the other hand, is mostly kept ionized and heated by the the cosmic UV background (UVB). It can be traced through Lyman alpha absorption ~\citep[e.g.,][]{art:abs2}, Lyman alpha emission \citep{art:silva16} or through HI 21 cm emission \citep[][hereafter K17]{art:takeuchi14,art:kooistra17} if the neutral fraction is high enough.\\

\noindent In \citetalias{art:kooistra17}, using a simple model based on the dark matter density field, it was shown that multiple current and upcoming radio telescopes have the sensitivity to possibly detect the HI 21 cm signal from the cold component in strong large-scale filaments at $z$ = 0.1 within $\sim100$ h integrations and with signal-to-noise ratio (S/N) of $\sim1-10$ for phase 2 of the Square Kilometre Array (SKA2). Such an observation could have the potential to provide an unbiased tracer of the underlying dark matter distribution ~\citep{art:cui18}.\\

\noindent The differential brightness temperature signal is proportional to the neutral hydrogen number density $n_{\rm HI}$ in cgs units following \citep{art:furl06}

\begin{align}
\delta T_{\mathrm{b}}^{\rm HI} (z) = & 5.48\times10^{-14}\times\frac{n_{\rm HI}(z)}{\left(1+z\right)H(z)}\\ \notag
&\times\left(1-\frac{T_{\rm CMB}(z)}{T_s}\right)\left[1 + H(z)^{-1}dv_r/dr\right]^{-1} \mathrm{K},\label{eq_c4:dtb}
\end{align}

\noindent where $H(z)$ is the Hubble parameter, $T_{\rm CMB}$ the temperature of the Cosmic Microwave Background (CMB), $T_{\rm s}$ the spin temperature of the gas and $dv_r/dr$ is the comoving gradient of the comoving velocity along the line of sight. The signal can be inferred from the gas density (including clumping) and temperature assuming an ionizing background radiation model. In \citetalias{art:kooistra17}, these quantities were determined from the density field of a dark matter (DM) only simulation by assuming that the baryonic matter follows the DM density field perfectly. From there, the ionization and neutral fractions and the temperature were derived in thermal and ionization equilibrium. The HI emission in that case is completely governed by the UVB, the density field and the cosmology.\\

\noindent However, we know from observations that filaments contain galaxies and quasars that provide an extra local source of heating and ionization. This, together with shock-heating negatively impacts the HI 21 cm signal. Therefore, for this study, we adopt a more sophisticated hydrodynamical simulation that includes both these effects, resulting in more realistic conditions in the IGM. This will allow us to better estimate the strength of the filament signal. We furthermore apply a strategy to find filaments that is applicable to both the simulations and the observations. We focus on the most sensitive of the upcoming radio telescopes and carry out a detailed calculation of the prospects for making a detection using either the first phase of the SKA (SKA1-mid) or the future upgrade to SKA2.\\

\noindent We begin this paper by describing the simulation and the framework we use to obtain a realistic filament signal in Section \ref{sec:sims}. The strategy that will be adopted for observations is laid out in Section \ref{sec:obsstrat}. Then estimates for observations with SKA1-mid and SKA2 will be made in Section \ref{sec:snr}. Finally, some additional observational effects will be discussed in Section \ref{sec:interfero}.  Throughout the paper we adopt the \citet{art:planck14} cosmological parameters, which are consistent with the latest ~\citet{art:planck18para} parameters within the errors.

\section{The HI 21cm signal from simulations}\label{sec:sims}
\noindent In this section we provide a description of the simulation that is used to determine a realistic filament signal and the method that was adopted to extract filaments from the box. We then discuss the major sources of uncertainty for such an estimate.

\subsection{The EAGLE simulation box}\label{sec:simbox}
\noindent The Evolution and Assembly of GaLaxies and their Environments (EAGLE) suite of simulations includes heating due to shocks as well as radiative transfer calculations to realistically propagate the photons from local galaxies into their surrounding environment \citep{art:eagle,art:eagle2}. We make use of the largest box that is available in the public data release \citep{art:eagle_pub2}. The box contains $1504^3$ dark matter particles with masses of $6.57\times10^6 h^{-1}\mathrm{M_\odot}$ and initially the same amount of baryonic particles with masses of $1.23\times10^6\, h^{-1}\mathrm{M_\odot}$. The volume of the simulation is $100\,\mathrm{Mpc}^3$.\\

\noindent The dataset provides us with the smooth particle hydrodynamical (SPH) particle data, namely the gas density, gas temperature and smoothing length of the particles, as well as the positions of halos. We calculate the HI properties of the gas, including the 21 cm differential brightness temperature signal directly on the SPH particles. In order to estimate the hydrogen neutral fraction in each cell, we assume ionization-recombination equilibrium and use the temperature directly from the simulation. The ionization fraction additionally depends on the electron density and the strength of the photo-ionizing background. We note that in this simulation, for starforming particles the temperature represents a parameterization of the effective pressure of the multiphase interstellar medium. Their temperature should therefore not be used to compute the ionization fraction. Since these particles will mostly be located within galaxies, we find that for the IGM in large-scale filaments considered in this work, both removing the starforming particles or fixing their neutral neutral fraction to unity yields the same result.\\

\noindent The HI 21 cm signal can then further be determined following the prescription outlined in \citetalias{art:kooistra17}, where we now use the case A hydrogen recombination rate and the corresponding fraction of Lyman $\alpha$ photons per recombination. Case A, where recombinations directly into the ground state are included, is more applicable to the gas considered here than case B, the latter of which was adopted in \citetalias{art:kooistra17}, since the IGM is optically thin \citep{book:draine11}. We then use the $YT$ package \citep{art:yt} to deposit the particles onto a regular grid with $1024^3$ cells, corresponding to a resolution of 66 $h^{-1}$kpc. We adopt a cubic SPH kernel and 58 nearest neighbors, similar to the kernel that was used in EAGLE itself \citep{art:eagle}. Further analysis in this paper is performed on this gridded box.\\

\noindent The photo-ionization rate is one of the main uncertainties in this estimation. Here, we consider three different models for the UVB: The \citet[][HM01 after this]{art:hm01} UVB gives the highest photo-ionization rate at $z$=0, whereas the \citet[][HM12 hereupon]{art:hm12} results in the lowest photo-ionization rate. These models adopt the same methodology, but the \citetalias{art:hm12} uses an updated description of the ionizing sources to adress the observational constraints available at the time. This includes X-ray emission from AGN and UV emission from starforming galaxies at all redshifts, as well as a more detailed treatment of absorption. The most recent continuation of this set of models is given by \citet[][henceforth P18]{art:hm18} with an intermediate photo-ionization rate and includes new constraints on the column density distribution of HI absorbers and a new treatment of the opacity for ionizing photons in the IGM. Each of these backgrounds will result in a different filament signal strength and this uncertainty in the intensity of the UVB and its implications will be discussed in more detail in Section \ref{sec:uvb}. We note that the EAGLE simulation itself adopts the \citetalias{art:hm01} UVB model. Adopting a different UVB photoionization rate for determining the HI fraction is not fully self-consistent, since the IGM temperature was computed assuming the \citetalias{art:hm01} model. Given that the two other UVB models that we consider have lower photoionization and heating rates, we can infer that by using the temperature provided by the EAGLE simulation, the HI fractions determined for the other UVB models are underestimated. This also results in a lower HI 21 cm signal. The estimates of the filament signal in this work that adopt the \citetalias{art:hm12} and \citetalias{art:hm18} UVB models should thus be considered as lower limits.\\

\subsection{Contamination by galaxies}\label{sec:galcont}
\noindent One thing to take into account for the observations is that emission from galaxies will contaminate the IGM signal. The voxels in the datacube containing the galaxies can be masked and the SKA has sufficient resolution to do this without losing significant fractions of the volume of a filament. The difficulty lies in finding the positions of the galaxies. Surveys, such as SDSS can provide the positions of the most massive galaxies, but that still leaves contamination by the weaker ones. In \citetalias{art:kooistra17}, this remaining contamination was estimated to be of the order of $\sim10\%$ after masking SDSS galaxies. Since the IGM signal estimates obtained from the EAGLE simulation used in this work are lower, signifying a lower neutral fraction and thus less HI gass, the contamination due to the faint galaxies not detected by SDSS will also be higher. However, SKA itself will be much more sensitive than SDSS, allowing the localisation and masking of faint galaxies well into the dwarf regime \citepalias{art:kooistra17}. Additionally, even if it will not be possible to mask all of the faint galaxies, the integrated signal would still contain emission from gas that has not been detected before and will therefore still be worth studying.\\

\begin{figure}
\centering
\includegraphics[width=0.49\textwidth]{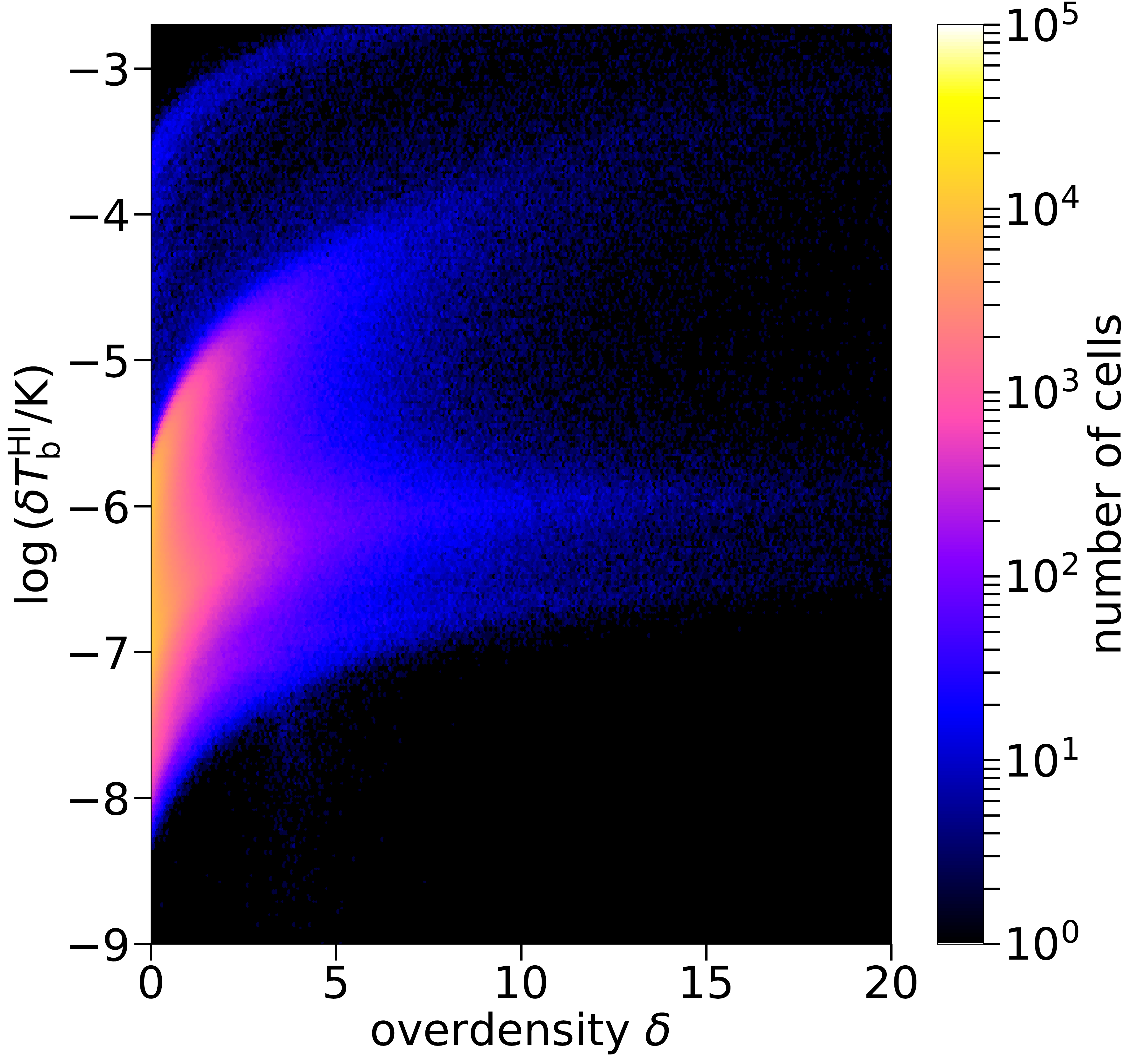}
\caption{Distribution of the HI 21 cm differential brightness temperature signal as a function of overdensity for cells in the EAGLE simulation, assuming the \citetalias{art:hm01} UV background and after masking cells with a radius of 100 $h^{-1}$kpc around the position of the halos. The cells at the top of the figure, above the main distribution are cells belonging to halos that were not completely masked. The colorbar shows the number of cells.}\label{fig:boxsigdens}
\end{figure}

\begin{figure*}
\centering
\includegraphics[width=\textwidth]{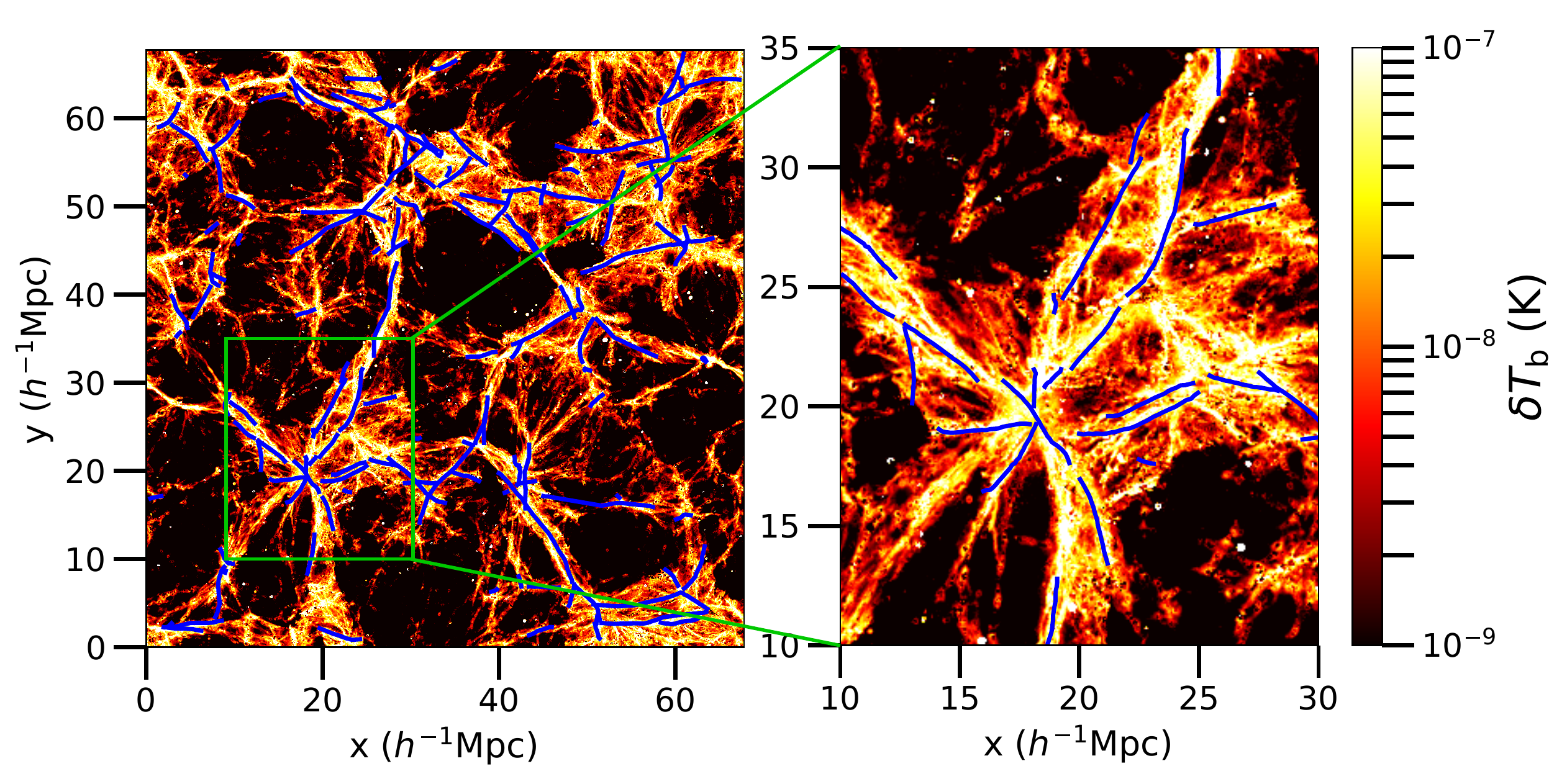}
\caption{Mean HI 21 cm brightness temperature in a slice of 5.3~$h^{-1}$Mpc of the EAGLE simulation for HI fractions based on the \citetalias{art:hm01} UVB. The blue lines denote the filament spines that were determined through the Bisous model code with a length of $l\geq5~h^{-1}$Mpc. The filaments are determined, based on the sample of star forming galaxies in the simulation with $M_{\rm gal}\geq 10^8M_{\odot}$. The right panel shows a zoom in of the region highlighted in the green box in the left panel. Cells around the position of galaxies were masked out to a radius of 100 $h^{-1}$kpc in these maps.}\label{fig:boxslice}
\end{figure*}

\noindent Galaxies and their circumgalactic medium (CGM) contain large amounts of neutral gas, but since we are targetting the HI gas in the IGM, we remove their signal by masking them with a fixed masking radius of 100 $h^{-1}$kpc. We explored the impact of adopting a different radius on the signal and found this radius to effectively remove most of the unwanted signal. Nonetheless, a small amount of contamination still remains from halos that are larger. Properly removing this contamination observationally would require an accurate measurement of the size of each galaxy. Due to the interferometric nature of the SKA, sources such as galaxies that are brighter than the extended emission from the IGM will also add strong sidelobe noise that needs to be removed from the datacube before extracting the IGM signal. This will be discussed in more detail in Section \ref{sec:interfero}.\\

\noindent Fig.~\ref{fig:boxsigdens} shows the distribution of HI 21 cm differential brightness temperature as a function of the overdensity of the cells in the simulation box, assuming the \citetalias{art:hm01} UVB and after masking the galaxies and their CGM. The baryonic physics included in EAGLE yield a wide distribution of signals for a given density that spans more than 4 orders of magnitude. The small number of bright cells that can be seen above the main distribution corresponds to regions in the CGM of large galaxies that were not masked with the adopted fixed masking radius. Since the number density of these cells is small, their effect on the targeted IGM signal is minimal.

\subsection{Filament extraction}\label{sec:simfil}
\noindent Although filaments can be easily identified by eye in slices of the simulation box, fully extracting these three-dimensional structures is difficult. For this study, we chose to use the Bisous model code by \citet{art:tempel14,art:tempel16} to find filaments in the EAGLE simulation. It has also been widely applied to observations. The Bisous algorithm models the three-dimensional structures in the distribution of the galaxies through a marked point process and only requires the galaxy positions as input. The statistical nature of the inferred filaments means that there is a significant probability that some of the identified filaments are not real, since a group of galaxies that formed along a line by chance, but whose underlying density field is not connected by a filament could still be inferred by the code as a connected filament. To minimize this effect in our sample, we only consider filaments that are longer than 5~$h^{-1}$Mpc. Moreover, the spatial distribution of the filaments should be closely connected to large galaxies. Therefore, we limit the sample to galaxies with masses above $10^8M_{\odot}$ in order to trace the stronger filaments, as well as those having non-zero starformation rate in EAGLE since filament galaxies can accrete gas from the filament and can thus be expected to have some ongoing starformation. The mean brightness temperature of HI 21 cm emission in a slice of the simulation box of width 5.3 $h^{-1}$Mpc can be seen in Fig.~\ref{fig:boxslice}, where the blue lines denote the inferred filament spines. Most of the structures in the box are traced well by the Bisous filaments.\\

\noindent The Bisous model code only provides the three-dimensional positions of the points defining the filament spine. Along the spine in the simulation, the width of a filament can vary. However, when dealing with observations, this information from the underlying density field is not available and so an assumption for the width needs to be made a priori. In this case we assume the filaments to have a radius of 0.5~$h^{-1}$Mpc, as was used in \citet{art:tempel14}, and mask all the cells that are at a distance greater than this radius away from the filament points to extract the cells in filaments. The signal of the complete filament is then given by the mean of all the cells that fall within the filament radius. Choosing a slightly smaller radius (i.e., 0.25 $h^{-1}$Mpc) did not significantly affect our mean signal estimates, but it gives more scatter in the signal, since some of the inferred spines can be misaligned with parts of the underlying density field. A filament radius of 1~$h^{-1}$Mpc does result in a slightly lower mean signal (i.e. $\sim$10-50$\%$), since some filaments are less wide and thus empty regions around the filaments will be included in the integration. The scatter for this large radius does decrease, since then most of the underlying density field of the filaments will be encapsulated by the cylinders around the filament spines, even if there is a slight misalignment. Because a larger physical radius also results in a larger area on the sky, we take the radius of 0.5~$h^{-1}$Mpc to avoid having to integrate over too large angular scales.\\

\begin{figure}
\centering
\includegraphics[width=0.48\textwidth]{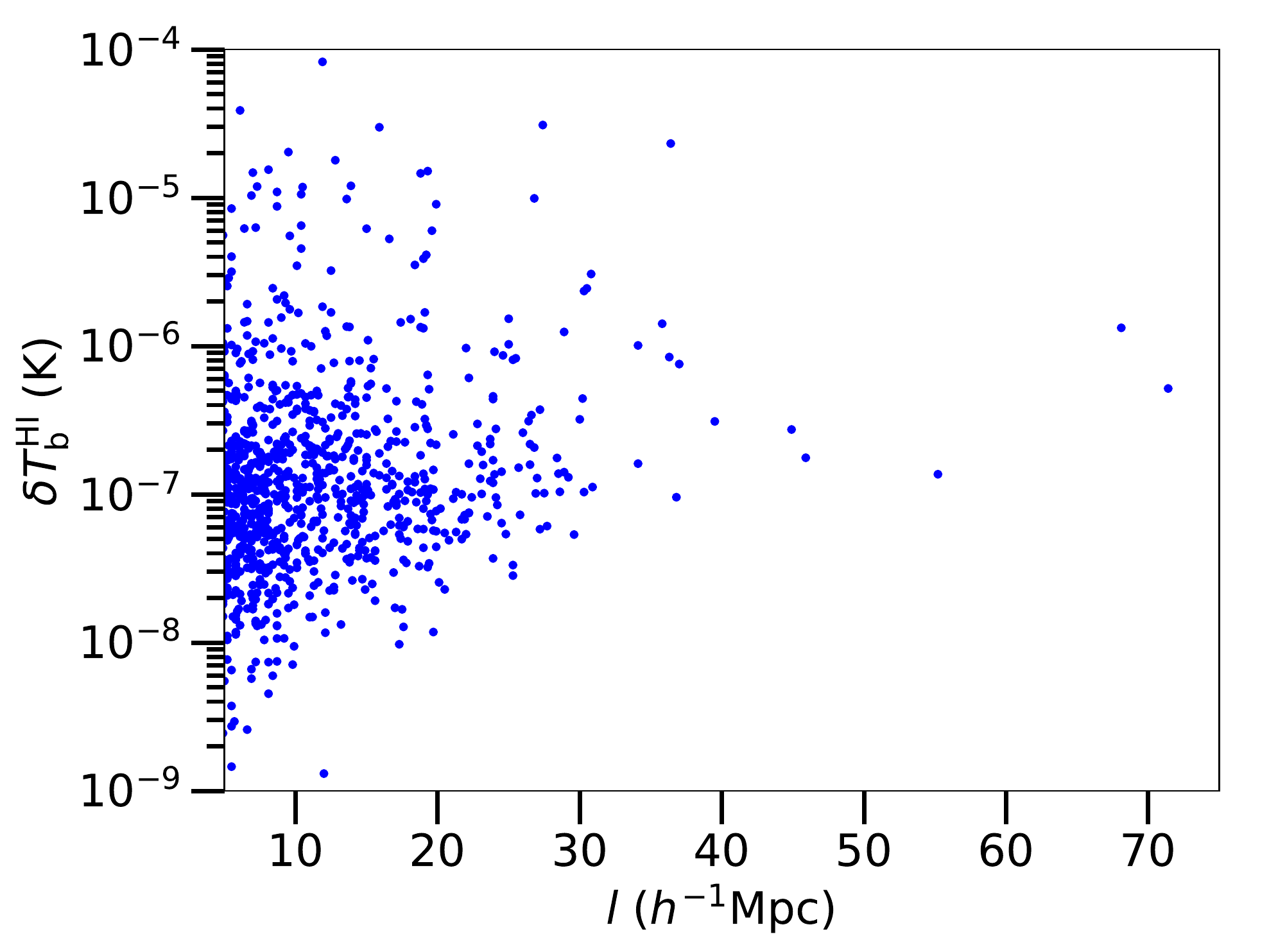}\\
\caption{Distribution of the mean signal per cell in filaments as a function of their length for Bisous filaments longer than 5~$h^{-1}$Mpc in the EAGLE simulation box. A filament radius of 0.5~$h^{-1}$Mpc was adopted together with the \citetalias{art:hm01} UVB.}\label{fig:filsigs}
\end{figure}

\noindent The distribution of the mean filament signal as a function of the filament length is shown in Fig. \ref{fig:filsigs}. The plot shows that there is a large range of mean signals for these filaments due to the different environments inside them. Howver, most of the low density gas sits at brightness temperatures around the median value of a few times $\sim10^{-7}$ K in this distribution. This corresponds to an HI column density of $\sim2-5\times10^{13}\,\mathrm{cm}^{-2}$ for a 100 km $\mathrm{s}^{-1}$ velocity width. In the inner most dense parts of the filaments they tend to be brighter and some dense clumps of gas can additionally drive up the mean filament signal to the higher outliers shown in Fig. \ref{fig:filsigs}. The outliers to lower mean signals are due to misalignments of the inferred filament spines with the underlying density distribution.\\ 

\noindent We note that many other methods have been developed to identify the large-scale structure components in simulations \citep[see e.g.,][for an overview of a number of different available codes]{art:libeskind17}. For observations, the large-scale structure is only traced by the positions of the galaxies. Recovering the filaments requires a method that can properly infer the structures based solely on the limited data from galaxy surveys. This makes the Bisous model a good choice for this study, since it can be applied directly to both the simulations and real data from galaxy redshift surveys. The same model will therefore come back in our observational strategy in Section \ref{sec:obsstrat}.\\

\noindent Another thing to note is that observations do not measure distances in physical distances. Instead, galaxy surveys observe in redshift-space, where the peculiar velocities of galaxies can add Doppler shifts to the measured redshifts \citep[e.g.,][]{art:davispeebles83,art:kaiser87}. This effect can result in errors in the inferred spatial positions of galaxies, whereby certain large-scale structures can appear to be more elongated along the line of sight than they are in reality. Therefore, before a filament finder can be applied to a sample of galaxies from a galaxy redshift survey, the redshifts first need to be converted to proper distances, which need to be corrected for the redshift-space distortions. The majority of the effects of velocities on redshift measurents can be supressed \citep[e.g.,][]{art:tegmark04,art:tempel14}, but for some filaments errors in the locations of the inferred spines can remain.

\subsection{UV background uncertainty }\label{sec:uvb}
\noindent As was briefly touched upon in Section \ref{sec:simbox}, another large uncertainty to the expected signal strength is the intensity of the UVB. For this reason, we considered three observationally driven models for the UVB. Although there is a large spread in the possible filament signal values, we use a mean value from the distribution shown in Fig. \ref{fig:filsigs} calculated for each UVB to determine its effect on the overall distribution. Since some of the highest signals are likely contaminated by remaining contribution from galaxies, we remove the 10$\%$ highest and lowest signal filaments to determine the mean. The resulting values are given in Table \ref{tab:uvbsigs}, together with the standard deviation of the cut distribution. For reference, we also give the corresponding HI column density over a 100 km $\mathrm{s}^{-1}$ velocity width. These values are consistent with the lower-density gas found through HI absorption measurements at low redshift \citep[e.g.,][]{art:teppergarcia12}. The last column in Table \ref{tab:uvbsigs} presents the column density based on the mean signal in the brightest 10 per cent of the cells in filaments in the simulation box. This shows that a significant fraction of the IGM gas in the filaments has densities one or two orders of magnitude higher than the mean and can thus also result in a stronger signal.\\

\begin{table} 
\caption{Estimated mean filament signals $<\delta T_{\rm b}^{\rm HI}>_{\rm fil}$ of the Bisous filaments for the three different UVB models (after removing the 10$\%$ lowest and highest outliers). The values are given in order of increasing HI photo-ionization rate ($\Gamma_{\rm HI}$) at $z$=0. The error on the signal denotes the standard deviation of the distribution of filament signals. The HI column densities $N_{\rm HI}^{\rm all}$ have been determined from the mean filament signals over a 100 km $\mathrm{s}^{-1}$ velocity width, corresponding to twice the filament radius of ~{$0.5\, h^{-1}\, \mathrm{Mpc}$} at $z$ = 0.01. $N_{\rm HI}^{10 \%}$ denotes the column density based on the mean filament signal in the brightest 10 per cent of the cells within filaments in the simulation.}
\renewcommand{\arraystretch}{1.2}
\begin{center}
\begin{tabular}{l c c c c}
 \hline
 \hline
  UVB model & $\Gamma_{\rm HI}$ at $z$=0 & $<\delta T_{\rm b}^{\rm HI}>_{\rm fil}$ & $N_{\rm HI}^{\rm all}$ & $N_{\rm HI}^{10 \%}$\\
  & ($10^{-14}\,\mathrm{s}^{-1}$) &  ($10^{-7}$ K) & (cm$^-2$) & (cm$^-2$)\\
 \hline
 \citetalias{art:hm12} & 2.3 & $6\pm6$ & $1\times10^{14}$ & $2\times10^{16}$\\ 
\citetalias{art:hm18} & 6.1 & $2\pm2$ & $4\times10^{13}$ & $8\times10^{15}$\\
\citetalias{art:hm01} & 8.4 & $2\pm2$& $3\times10^{13}$ & $5\times10^{15}$\\
 \hline
\end{tabular}
\end{center}
\label{tab:uvbsigs}
\end{table}

\noindent Due to its high photo-ionization rate, the \citetalias{art:hm01} model also gives the lowest signal limit.  However, the photo-ionization rate of this model lies above the limits derived from observations of H$\alpha$ emission in a nearby galaxy \citep{art:uvbmeasure}. The \citetalias{art:hm12} model on the other hand, results in the highest signal, but its photo-ionization rate lies below the observed limits. The UVB model by \citetalias{art:hm18} is fully consistent with those limits and results in a filament signal that is a factor $\sim$3 lower than the one derived using \citetalias{art:hm12}. The difference between the mean signals from the \citetalias{art:hm01} and \citetalias{art:hm18} is negligible within the standard deviations of the distributions. We assume these values to be representative for typical filaments and we note that the difference in the resulting filament signal strengths for the two extreme models is only a factor of 3. For further estimates of the detectability, we consider both extreme cases: $\delta T_{\rm b} = 2\times10^{-7}$ K from \citetalias{art:hm01} as the lower limit and $\delta T_{\rm b} = 6\times10^{-7}$ K assuming \citetalias{art:hm12} as the upper limit.

\section{Observational strategy}\label{sec:obsstrat}
In order to maximize the chance of a detection, it is essential to first determine where on the sky the best candidate filaments can be found. The only resource available to find target filaments are the positions of the galaxies detected beforehand. In the same way that we determined filament signals from the galaxy catalogue of the simulation in Section \ref{sec:simfil}, we propose to use galaxy surveys to first determine the locations of the filament spines and then integrate the emission around these spines to obtain a signal.\\

\noindent In this Section we apply this strategy to existing galaxy surveys to determine how many filaments are accesible to the SKA. We then use the average filament signals obtained in Section \ref{sec:simfil} to estimate the S/N for the integrated signal per filament with the SKA, where the S/N then depends on the three-dimensional orientation of the filament. Here, we outline which steps will be required to make a detection and how we estimate the noise that can be expected for such an observation with the SKA.\\

\noindent In this case we consider the two phases of the telescope separately. The first phase of the SKA will consist of the 64 13.5m dishes currently operating as the Karoo Array Telescope (MeerKAT) together with 133 dishes of 15m diameter that will be added to it\footnote{See Baseline Design Document version 2 at: \href{https://www.skatelescope.org/key-documents/}{https://www.skatelescope.org/key-documents/}\label{footnote:ska}}. For its second phase, this system will be significantly expanded to a total of 1500 dishes, which will result in an unprecedented sensitivity at these frequencies. The relevant properties of these two arrays that will be used in this study are summarized in Table \ref{tab:specs}.\\

\begin{table} 
\caption{Properties of the two phases of the SKA (see footnote \ref{footnote:ska}).}
\renewcommand{\arraystretch}{1.2}
\begin{center}
\begin{tabular}{l c c}
 \hline
 \hline
 Parameter & SKA1-Mid & SKA-2\\
 \hline
 Number of dishes, $N_{\rm dish}$ & 197 & 1500 \\
 Dish diameter, $D_{\rm dish}$ (m) & 64$\times$13.5 + 133$\times$15 & 15\\
 Total collecting area, $A_{\rm tot}$ ($\rm{m}^2$) & 32,664 & 265,071\\
 System temperature, $T_{\rm sys}$ (K) & 20 & 30\\
 Aperture efficiency, $\epsilon_{\rm ap}$ & 0.8 & 0.8\\
 Field of View ($\rm{deg}^2$) & $\sim$1 & $\sim$1 \\
 Angular resolution (arcsec) & $\sim$0.3 & $\sim$0.1\\
 System Equivalent & &\\
 Flux Density, SEFD (Jy) & 2.1 & 0.4 \\
 \hline
\end{tabular}
\end{center}
\label{tab:specs}
\end{table}

\subsection{Filament catalogues from SDSS, 2MRS and 6dF}\label{sec:filcat}
Given their high sensitivity, both SKA1-mid and SKA2 will be able to detect extremely faint galaxies and therefore it would be possible to trace almost the complete cosmic web within their surveys. However, until data from such surveys becomes available, it is necessary to resort to existing galaxy surveys in order to determine the location of filament spines on the sky. In this case, we use the galaxy catalogues from three large-area galaxy redshift surveys to identify realistic filaments as a proxy for what the SKA should be able to detect in its survey volumes.\\

\noindent One of the largest and most dense samples of galaxy redshifts was obtained by SDSS. We adopt the filament catalogue that was described in \citet{art:tempel14}, whose data is publicly available. The catalogue uses a sample of 499340 galaxies in the redshift range 0.009 $\leq z \leq$ 0.155 and is limited in magnitude by the spectroscopic sample \citep{art:ssdssspecsamp}.\\

\noindent Additionally, we applied the Bisous model to the galaxy sample of the 2MRS survey and to a combined sample of 2MRS and 6dF galaxies. The 6dF sample of 126754 sources is limited in magnitudes by $m_{\rm K} \leq$ 12.65, $m_{\rm H} \leq$ 12.95, $m_{\rm J} \leq$ 13.75, $m_{\rm r_{\rm F}} \leq$ 15.60 and $m_{\rm b_{\rm J}} \leq$ 16.75, and has a median redshift of 0.053 \citep{art:6df3}. In the case of 2MRS, the sample contains 44599 galaxies and is limited in magnitude by $m_{\rm K_{\rm s}} \leq 11.75$. The survey only probes out to redshifts of $z \approx$ 0.05, but it covers a larger area of the sky than the other two surveys \citep{art:2mrs}. Because the 2MRS galaxies are used in both filament catalogues, there will be some overlap between the filament spines derived from them.\\

\subsection{Noise estimation}\label{sec:noise}
The signal-to-noise ratio of a filament depends on both the signal itself and on the noise originating from the instrument and the survey characteristics. We assume every filament has the same signal, so that the S/N only varies depending on the filament size on the sky and the telescope sensitivity.\\

\begin{figure}
\centering
\includegraphics[width=0.49\textwidth]{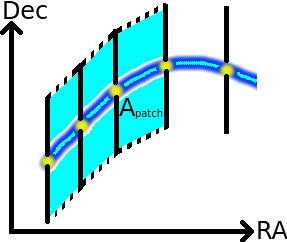}
\caption{Sketch of the filament noise integration scheme. The blue line shows the spine of the filament in the sky plane, where the yellow dots denote the filament points from the filament catalogue. The black lines denote the angular size of the 1 $h^{-1}$Mpc filament diameter at each point, $\Theta_i$. The cyan shaded trapezoid areas ($A_{\rm patch}$) between every point are added together in a weighted sum, according to Equation \ref{eq:totnoise}.}\label{fig:filnoise}
\end{figure}

\noindent For the integration of the signal in real data, the filament would be split up into resolution elements, whose angular size corresponds to the filament diameter of $0.5~h^{-1}$Mpc. However, since the resolution of the observation itself is higher, the noise will add together depending on the size of such an integration element. The depth of the filament within that element then sets the frequency range for the integration. In order to estimate the noise for each filament, we take every point of the spine given by the filament catalogue, then use its redshift to determine the angular size of the filament diameter and add together the trapezoid shaped patches set by two neighboring spine points. The parallel sides of the trapezoid are set by the angular size by going up and down 0.5~ $h^{-1}$Mpc in declination, whereas the other two sides of the trapezoid are formed by connecting the tops and bottoms of the parallel sides. Fig.~\ref{fig:filnoise} gives a sketch of how a filament is divided up into the trapezoid patches along the sky for the integration.\\

\noindent The area $A_{\rm patch}$ on the sky of a single patch is set by the angular size $\Theta$ as:
\begin{equation}
 A_{\rm patch} = \frac{\Theta_i + \Theta_{i+1}}{2}\times\Delta_{\mathrm{RA}},
\end{equation}
where $\Theta_i$ is the angular size at point i, $\Theta_{i+1}$ the angular size at the adjacent point and $\Delta_{\rm RA}$ is the difference in right ascension between the two points. 
The noise of a radio telescope is given by
\begin{equation}
\Delta T^{\rm N} = \frac{\lambda_0^2(1+z)^2}{\Delta\theta^2\epsilon_{\rm ap}A_{\rm tot}}\frac{T_{\rm sys}}{\sqrt{\Delta\nu t_{\rm obs}}},\label{eq:telnoise}
\end{equation}
where the rest frame wavelength of the observed line is denoted by $\lambda_0$ and $\Delta\theta$ is the angular resolution. As an approximation, we set $\Delta\theta^2=A_{\rm patch}$. The parameter $\epsilon_{\rm ap}$ denotes the aperture efficiency, $A_{\rm tot}$ is the total collecting area of the telescope, $T_{\rm sys}$ the system temperature, $\delta\nu$ the frequency bandwidth over which is integrated for the observations and $t_{\rm obs}$ is the integration time \citep[][]{art:furl06,art:kooistra17}. Equation \ref{eq:telnoise} implies that the sensitivity will be higher when the filament is orientated perpendicular to the line-of-sight. This results in a noise level of 5.2$\times10^{-6}$ K (9.6$\times10^{-7}$ K) for SKA1-mid (SKA2) for an angular resolution of 10 arcmin, and frequency resolution of 20 kHz with an integration time of 120 h at $z = 0.01$.\\

\noindent All the elements along the filament spine are then added together in the following manner to obtain the total noise per filament:
\begin{equation}
 \sigma_{\rm fil} = \frac{1}{n_{\rm patch}}\sqrt{\sum_i\sigma_{\mathrm{patch},i}^2},\label{eq:totnoise}
\end{equation}
with $n_{\rm patch}$ denoting the number of patches.\\ 

\noindent The angular size of very local filaments can be quite large. If the scales of the fluctuations in the IGM become too large, the spatial filtering of the telescope would cause a reduction in the S/N and a deviation from Equation \ref{eq:telnoise}. Therefore, we limit the filaments considered here to a minimum redshift of $z=0.01$. The effect of the large angular scale of the nearby filaments will be discussed in more detail in Section \ref{sec:deconv}. We note that for the observations the voxels containing emission from galaxies would have to be masked, reducing the volume that is integrated over. However, since the size of the galaxies is small compared to the size of a filament, this would require only a minor correction and is not included here.

\begin{table*} 
\caption{Properties and predicted SKA1-mid and SKA2 signal-to-noise values of the ten best filaments in each of the three galaxy catalogues with an integration time of 120 h for the \citetalias{art:hm12} upper limit signal of $\delta T_{\rm b} = 6\times10^{-7}$ K and the \citetalias{art:hm01} lower limit of $\delta T_{\rm b} = 2\times10^{-7}$ K. The first column in the table shows catalogue the filaments belong to, the second gives the ID number of the filament and third shows the redshift of the filament. The angular size corresponding to the filament diameter at redshift $z$ is given by $\Theta(z)$ and its length by $l$. The remaining columns give the upper and lower S/N values for both SKA phases.}
\renewcommand{\arraystretch}{1.2}
\begin{center}
\begin{tabular}{l c c c c c c c c}
 \hline
 \hline
 Filament & ID  & $z$ & $\Theta(z)$ & $l$ & S/N  & S/N & S/N & S/N\\
 Catalogue & & & (deg) & ($h^{-1}$Mpc) & SKA1-mid & SKA2 & SKA1-mid & SKA2\\ 
 &  & & &  & HM12 & HM12 & HM01 & HM01\\
 \hline
 SDSS & 1 & 0.01 & 1.9 & 10.6 & 33 & 182 & 11 & 61\\
  &  2 & 0.011 & 1.7 & 10.1 & 29 & 158 & 9.5 & 53\\
  &  3 & 0.011 & 1.7 & 7.3 & 26 & 140 & 8.7 & 47\\
  &  4 & 0.013 & 1.5 & 20.6 & 25 & 133 & 8.3 & 44\\
  &  5 & 0.012 & 1.6 & 15.6 & 24 & 130 & 8.0 & 43\\
  &  6 & 0.012 & 1.6 & 7 & 20 & 109 & 6.7 & 36\\
  &  7 & 0.014 & 1.4 & 18.6 & 19 & 100 & 6.3 & 33\\
  &  8 & 0.012 & 1.6 & 15.7 & 17 & 91 & 5.6 & 30\\
  &  9 & 0.013 & 1.5 & 15.6 & 17 & 91 & 5.6 & 30\\
  &  10 & 0.014 & 1.4 & 9.7 & 16 & 90 & 5.4 & 30\\ \hdashline
 2MRS & 11 & 0.011 & 1.7 & 11.5 & 90 & 500 & 30 & 167\\
  + & 12 & 0.012 & 1.6 & 7 & 40 & 214 & 13 & 71\\
  6dF & 13 & 0.012 & 1.6 & 19.5 & 40 & 214 & 13 & 71\\
  & 14 & 0.01 & 1.9 & 13 & 38 & 200 & 13 & 67\\
  & 15 & 0.014 & 1.4 & 8.5 & 38 & 200 & 13 & 67\\
   & 16 & 0.012 & 1.6 & 25.5 & 26 & 140 & 8.7 & 47\\
   & 17 & 0.017 & 1.1 & 19 & 23 & 128 & 7.7 & 43\\
   & 18 & 0.014 & 1.4 & 35 & 22 & 120 & 7.4 & 40\\
  & 19 & 0.013 & 1.5 & 24 & 21 & 113 & 6.9 & 38\\
  & 20 & 0.015 & 1.3 & 27.5 & 21 & 111 & 6.9 & 37\\ \hdashline
  2MRS & 21 & 0.01 & 1.9 & 16 & 80 & 429 & 27 & 143\\
  & 22 & 0.014 & 1.4 & 9.5 & 35 & 194 & 12 & 65\\
  & 23 & 0.01 & 1.9 & 9 & 35 & 188 & 12 & 63\\
  & 24 & 0.019 & 1.0 & 17 & 30 & 167 & 10 & 56\\
  & 25 & 0.025 & 0.8 & 9 & 29 & 158 & 9.5 & 53\\
  & 26 & 0.012 & 1.6 & 22 & 26 & 143 & 8.7 & 48\\
  & 27 & 0.011 & 1.7 & 18.5 & 25 & 133 & 8.3 & 44\\
  & 28 & 0.013 & 1.5 & 6.5 & 24 & 128 & 8.0 & 43\\
  & 29 & 0.014 & 1.4 & 40 & 24 & 128 & 8.0 & 43\\
  & 30 & 0.022 & 0.9 & 11.5 & 22 & 120 & 7.4 & 40\\
 \hline
\end{tabular}
\end{center}
\label{tab:snr}
\end{table*}

\section{SKA signal-to-noise predictions}\label{sec:snr}
Applying the methods described in Section \ref{sec:noise} to all the filaments in the filament catalogues described in Section \ref{sec:filcat} and dividing the signal estimate by the total noise then yields the expected S/N. Because of the large variety in HI photo-ionization rates for the UVB models discussed in Section \ref{sec:snr}, we determine the S/N for the two limits derived there: $\delta T_{\rm b} = 2\times10^{-7}$ K (lower) and $\delta T_{\rm b} = 6\times10^{-7}$ K (upper). We assume an integration time of 120 h for the observations.\\

\noindent The galaxy catalogues also cover areas of the sky that are not accesible to the SKA. Given its latitude of $\sim$-30$^\circ$, we remove all filaments from the sample that fall outside the declination range of +60$^\circ$ to -90$^\circ$. Out of the 9477, 6779 and 2603 filaments of length equal to or greater than 5~$h^{-1}$Mpc at $z\geq$0.01 for the SDSS, 2MRS+6dF and 2MRS filament catalogues, respectively, 85, 231 and 162 filaments can be detected with SKA1-mid at S/N$\geq$2, assuming the lower limit signal. In the case of the upper limit signal, the number of filaments increases to 475, 860 and 637, respectively. For SKA2, the S/N is $\sim5.5$ times higher, following from the difference in the collective areas using Equation \ref{eq:telnoise}, and thus many more filaments become available for individual detections. The ten highest S/N filaments from each catalogue and their properties are summarized in Table \ref{tab:snr}. The maximum integrated S/N value of a filament we estimate for SKA1-mid is 90, whereas the same filament has a S/N value of 500 with SKA2. Therefore, although SKA1-mid can make initial detections of some of the filaments by integrating along the filament, an instrument as sensitive as SKA2 would even be able to map out the brightest parts of the filaments. Looking at the highest S/N filament with SKA2 (filament 11 in Table \ref{tab:snr}), the signal would reach S/N = 2 with an integration time of only $\sim$7-62s, depending on the strength of the UVB. The S/N = 2 integration time for the same filament with SKA1-mid would be $\sim$4-32m.\\

\noindent We point out again that there is significant scatter in the filament signals, as shown in Fig. \ref{fig:filsigs}. Therefore, there will be filaments presented here that will be detected with an even higher S/N. Overall it can therefore be expected that a significant number of robust detections will be made already with SKA1-mid. The advantage of the integration method presented here is that, in principle, the integration can be performed with data from any galaxy survey with the same instruments, as long as the integration time per pointing of the survey is long enough.

\section{Effects of interferometers}\label{sec:interfero}
Radio interferometers, such as the SKA, are limited by their baselines in which scales they are sensitive to. For a point source, the sensitivity of the telescope corresponds to the value presented in Equation \ref{eq:telnoise}. For diffuse emission, the situation becomes more complicated due to spatial filtering, which could cause additional loss of signal on the larger scales. In the next section, we therefore make a rough estimate of the magnitude of this effect by considering two-dimensional images of a filament from the simulation.

\subsection{Spatial filtering}\label{sec:deconv}
The S/N estimates presented so far have assumed that the telescope can perfectly probe the complete filaments. However, due to the nature of an interferometer, only scales smaller than the scale corresponding to the shortest baseline will be resolved. For the SKA, the minimum baseline will be $\sim 20$ m, corresponding to an angular scale of $\sim$36 arcmin. Also, since the UV-plane is not fully sampled, there will be significant spatial filtering that will cause a signal loss on the more diffuse structures. In order to estimate this loss, we manually extracted a $\sim$10 $h^{-1}$Mpc filament from the simulation box. We then convolved an image of the filament with a point spread function (PSF) of SKA1-mid, where the angular scale of the pixels in the PSF image changes as a function of redshift. This way, we can directly compare between each redshift, since the same filament is imaged in all cases. The baseline design for SKA2 is not yet known, but since it will more dishes than SKA1-mid, as well as include the dishes already in SKA1-mid, we only perform the calculation for SKA1-mid and expect that the performance of SKA2 will be even better.\\

\noindent The PSF is calculated in two steps. First, a measurement set is created using the publicly available SIMMS package\footnote{Written by Sphesihle Makhathini: \href{https://github.com/radio-astro/simms}{https://github.com/radio-astro/simms}}. Here we adopt the antennae positions of SKA1-mid and choose a pointing to -30$^\circ$ in declination and 0h00m RA, at the frequency corresponding to the required redshift and for a single frequency channel of 20 kHz. In reality, one would have to integrate over multiple frequency channels to cover an entire filament, but here we treat it as if the filament was entirely in the plane of the sky to allow us to estimate the spatial filtering in the angular directions.\\
 
\noindent In the next, step we created an image of the PSF from the measurement set using the w-stacking clean imager \citep[WSClean;][]{art:wsclean}, which allows us to sample the PSF in any size and scale. We calculate a separate PSF for each redshift. Here, we used images with a size of $2048\times2048$ pixels, where the angular pixel scale depends on the redshift. We apply uniform weighting to the visibilities. This gives the noise level that was estimated with Equation \ref{eq:telnoise}, but it also results in the highest resolution. Other weighting schemes, such as robust or natural weighting will result in better surface brightness sensitivity. The angular scale of a pixel in this $2048\times2048$ pixel image then becomes 22.276$^{\prime\prime}$,11.164$^{\prime\prime}$, 4.497$^{\prime\prime}$ and 2.276$^{\prime\prime}$ at z = 0.01, 0.02, 0.05 and 0.1, respectively. The corresponding maximum baselines probed by this PSF then respectively are 2.0, 4.0, 10 and 21 km, which yields much lower resolution than the $\sim$0.3$^{\prime\prime}$ FWHM of the SKA beam. Therefore, we are under-sampling the PSF. However, in order to cover a field of view (FoV) that encompasses the entire filament, this was the maximum resolution we could manage computationally, given the required memory usage. In order to check the robustness of the PSF estimate, we recalculated the PSF by excluding all baselines that are longer than the scale equivalent to the size of a pixel. This resulted in PSFs identical to the ones without a baseline cut and therefore we believe that the estimates presented here are reasonable. Doing a more detailed calculation would require multiple pointings.\\

\noindent Using the SPH kernel we deposited the simulated filament onto a $2048\times2048$ pixel image, the same size as the PSF image. Finally, we convolved the simulated image with the PSFs at different redshifts to generate dirty images of the filament. We note that the outer regions of the PSF image are noisy and would result in strong edge effects in the dirty image that are not real. We therefore multiplied the PSF image with a Gaussian of width 410 pixels (1/5th the size of the image) before the convolution in order to mitigate this effect. Since the exercise here is meant to quantify the effect of the spatial filtering on the filamentary structure, no noise was added to the images.\\

\begin{figure*}
\centering
\includegraphics[width=\textwidth]{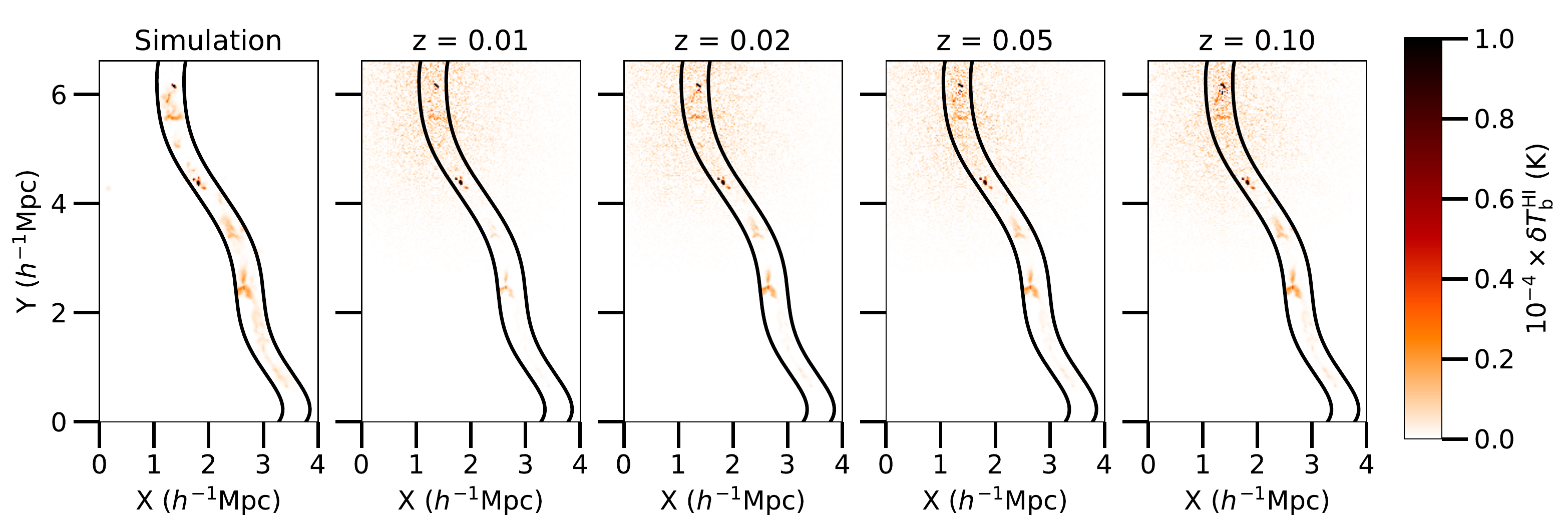}
\caption{Convolution and cleaning result of a simulated filament. SPH particles around the positions of galaxies have been masked out to a radius of 100 $h^{-1}$kpc. The most left image shows the filament directly from the simulation. The four other images show how the  filaments would appear observed at different redshifts by convolving the simulated image with the corresponding PSF of the beam and then cleaning it. In all images, the black lines highlight an area with a width of 1 $h^{-1}$Mpc around the interpolated spine of the filament. Significant sidelobe noise from the brightest peaks remains at the top of the image.}\label{fig:deconv}
\end{figure*}

\noindent The images were then cleaned using the \textit{deconvolve} task in the Common Astronomy Software Applications package \citep[CASA;][]{art:casa}, adopting the multiscale algorithm with a gain of 0.7 and 30,000 iterations. However, the deconvolution results were unstable due to bright sidelobes from the strong galaxies inside the filament outshining the IGM and being interpreted by $CASA$ as sources. This will also be an issue for observations, where galaxies will contaminate the signal in a similar way. We furthermore attempted the same exercise on the filament image where SPH particles within a radius of 100 $h^{-1}$kpc from the galaxy positions were removed before gridding the image. This image is shown in the furthest left panel of Fig. \ref{fig:deconv}. However, the cleaning algorithm still diverges on a solution and, as can be seen in the partially cleaned images in the remaining panels of Fig. \ref{fig:deconv} there is added noise from the brighter pixels at the top. The major structures of the filament are recovered well and the integrated filament signal between the black curves becomes lower from a factor $\sim$2 at z = 0.01, up to a factor of 1.1 at z = 0.1.  Therefore, we do not believe the spatial filtering to significantly affect the IGM emission. It would, however, require significant tweaking of the parameters to get the cleaning algorithm to converge and also recover the brighter structures at the top of the image with less noise. This falls outside the scope of this paper and we leave this for a future publication.\\

\begin{figure*}
\centering
\includegraphics[width=\textwidth]{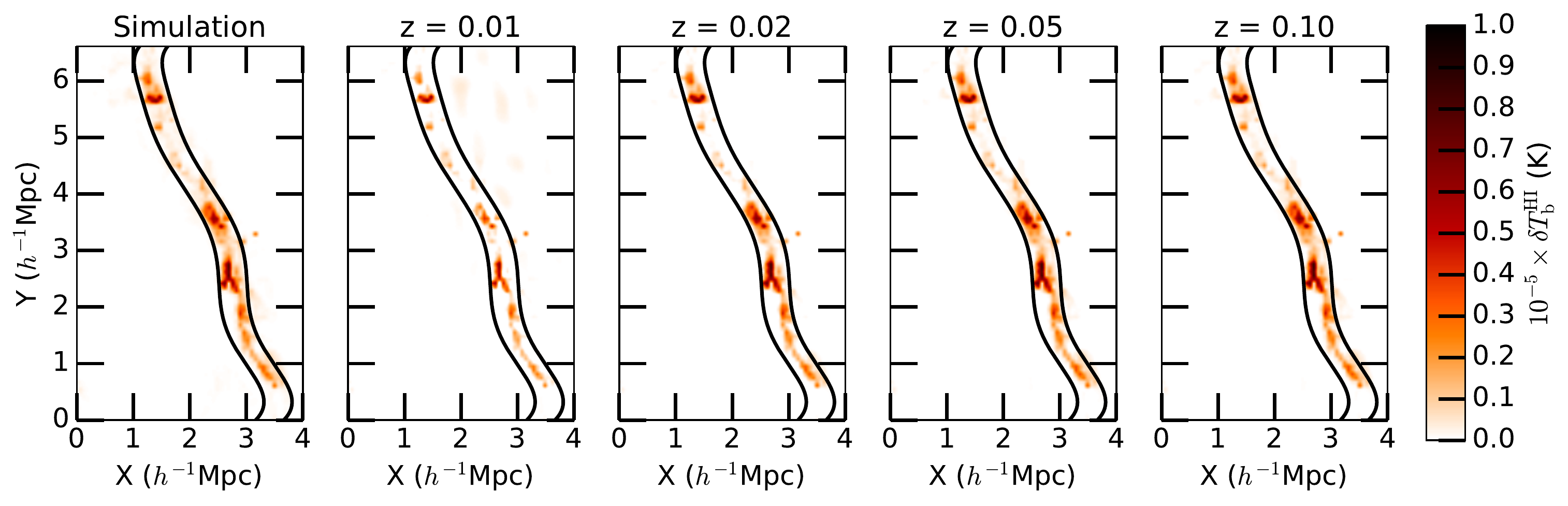}
\caption{Same as Fig. \ref{fig:deconv}, but now for a filament image that was smoothed to a resolution of 0.13 $h^{-1}$Mpc. The ratio of the signals within the black curves of the simulated image and the cleaned images is $\sim$3, 1.2, 1.0 and 1.0 for the images at z = 0.01, 0.02, 0.05 and 0.10, respectively.}\label{fig:deconv_smoothed}
\end{figure*}

\noindent Nonetheless, as an additional test we attempted the same cleaning exercise as described above, but then on a smoothed version of the filament image. In this case we extract the same filament with a resolution of 0.13 $h^{-1}$ Mpc and linearly interpolate it to the 2048x2048 pixel image size used for the PSF of SKA. Due to the interpolation, this removes the small scale fluctuations, which makes it easier for the cleaning algorithm to converge, but the spatial filtering would conversely be stronger. This image is shown in the left panel of Fig. \ref{fig:deconv_smoothed}. The cleaned images are given in the second and following panels. As can be expected, some of the diffuse emission is filtered away by the telescope in all cases. For the lowest redshift, the effect is strongest, since the physical size of the fluctuations corresponds to the largest angular scale on the sky. We determine the signal loss by adding all the cells in between the two black lines and dividing the value of the signal in the non-convolved image with that of the convolved images. This gives signals lower by a factor of $\sim$3 and $\sim$1.2 for $z$ = 0.01 and 0.02, respectively. For the $z$ = 0.05 and $z$ = 0.1, the signal loss is less than $\sim1\%$. Therefore, only for the nearest filaments does the spatial filtering become significant.\\

\noindent Given that both the real and simulated filaments contain structures at smaller scales, as shown in Fig. \ref{fig:deconv} already, the signal loss in reality should be lower than this. However, removing the bright contaminating sources will prove challenging for such an experiment and would require modeling the sources and removing them before imaging. Methods for this have already been developed for 21 cm Epoch of Reionization experiments \citep[e.g.,][]{art:sagecal}.

\section{Conclusion}\label{sec:concl}
Detecting the 21 cm signal from the neutral hydrogen gas in the IGM is very challenging. It requires sensitive telescopes to reach noise levels below the 21 cm signal from the small amount of neutral gas. Therefore, in this study, we determined the prospects for the detection of the integrated HI 21 cm signal of large-scale filaments with the most powerful upcoming radio telescopes, SKA1-mid and SKA2.\\

\noindent We made use of the density field and gas temperature from the EAGLE simulation in order to realistically estimate the HI 21 cm brightness temperature signal in the IGM. The Bisous filament finder code was then used to extract filaments from the simulation. We find, although there is significant scatter, a conservative estimate of the integrated mean filament signal of $2-6\times10^{-7}$ K, depending on the strength of the UVB.\\

\noindent We then took filaments from catalogues inferred from existing galaxy surveys to identify realistic filaments within the sky accessible to the SKA to estimate the S/N that can be expected with both SKA1-mid and SKA2. The signal was determined for three different estimates of the UVB. This study yields $\sim$478-1972 filaments that lie within the detection threshold of SKA1-mid with 120h integrations, where the strongest results in a S/N value of 30 (167) for SKA1-mid (SKA2), assuming the most pessimistic UVB and ignoring spatial filtering due to the array. The noise for this filament would already result in S/N = 2 with with integration times of $\sim$4-32m with SKA1-mid and $\sim$7-62s for SKA2.\\

\noindent In order to estimate the effect on this signal of observing with an interferometer, we made an estimation of the magnitude of the signal loss due to the spatial filtering. This showed that, for the closest filaments, the signal can decrease up to a factor of $\sim3$. We also find that bright sources, such as galaxies will have to be carefully modeled and removed in order to recover the IGM signal.\\

\noindent Therefore, SKA1-mid will be able to make initial detections of the integrated 21 cm signal of a large sample of filaments, whereas SKA2 will open up the possibility for large statistical studies of the filament signals, as well as potentially mapping parts of them. The direct detection of neutral gas from the IGM in large-scale filaments will greatly contribute to constraining ionization conditions within the IGM in the local Universe.

\section*{Acknowledgements}
RK, MS and SZ thank the Netherlands Foundation for Scientific Research for support through the VICI grant 639.043.006.\\
ET was supported by ETAg grants IUT40-2, IUT26-2 and by EU through the ERDF CoE grant TK133 and MOBTP86.\\
RK would like to thank Florent Mertens, Bharat Gehlot, Kyle Oman, Ed Elson and Mario G. Santos for the insightful discussions and help with learning to run some of the software. Our gratitude also goes out to Francesco Haardt and Roger Deane for supplying us with some of the required data and helping us find suitable software to use.\\
SZ acknowledges support by the Israel Science Foundation (grant no. 255/18).

\bibliographystyle{mnras}
\bibliography{skafils.bib}

\end{document}